\documentclass[10pt,a4paper]{article}
\usepackage{psfig}
\begin{document}
\textwidth=135mm
 \textheight=200mm
\begin{center}
{\bfseries 
Neutrino emissivities in 2SC color- superconducting quark matter}
\vskip 5mm
J. Berdermann$~^1$    
\vskip 5mm
{\small 
{\it $~^1$ Institut f\"ur Physik der Universit\"at, 
D-18051 Rostock, Germany}} 
\end{center}
\vskip 5mm

\centerline{\bf Abstract}{ 
The phase structure and equation of state for two-flavor quark matter 
under compact star constraints is studied within a nonlocal chiral quark model.
Chiral symmetry breaking leads to rather large, density dependent quark masses 
at the phase transition to quark matter. 
The influence of diquark pairing gaps and quark masses on density 
dependent emissivities for the direct URCA is discussed. 
Since $m_u>m_d$, the direct URCA process due to quark masses cannot occur.
We present cooling curves for model quark stars and discuss their relation to 
observational data.
\vskip 5mm
}

\section{\label{sec:intro}Introduction}{
Emissivities and mean free paths of photons and neutrinos are essential for 
most astrophysical phenomena (supernovae, neutron star cooling, gamma ray 
bursts(GRB), pulsar kicks \cite{Berdermann:2006rk} etc.).
Their correct treatment is one of the challenging tasks in astrophysics.
Theoretical predictions of quark matter properties inside compact stellar 
objects including the possibility of different color-superconducting phases, 
resulted in a  number of emissivity calculations in quark matter until now 
(e.g. \cite{Iwamoto:1982,Jaikumar:2005hy,Schmitt:2005wg,Anglani:2006br}).
The first calculation of neutrino emissivities in quark matter has been done 
by Iwamoto \cite{Iwamoto:1982}. He found that the matrix element of the direct 
URCA process would vanish and this important cooling process in quark stars 
could not occur if one neglects quark-quark interactions and quark masses.
Most calculations consider the effect of quark-quark interactions to obtain a 
finite matrix element since the current up and down quark masses are small and 
their influence on the emissivity is negligible. 
However, the up and down quark masses can be up to almost two orders of 
magnitude larger than the current quark mass in the vicinity of the phase 
transition to quark matter. 
This can be shown within NJL-type chiral quark models \cite{Buballa:2003qv}.   
The influence of both quark masses and diquark pairing on the direct URCA 
emissivities is discussed in the following sections. 
}
\section{Relativistic chiral quark model}
{
The grand canonical potential for quark matter in a 2SC superconductor is
{\small
\begin{eqnarray}
\Omega(\mu_B,\mu_Q,\mu_8,T)&=&
(1-\alpha)\frac{\phi_u^2+\phi_d^2}{8~G_S}+\alpha\frac{\phi_u \phi_d}{4~G_S}
+\frac{\Delta^2}{4~G_D}\nonumber\\
&&-2\int \frac{{\rm d}^3 p}{(2\pi)^3}\sum\limits_{a=1}^{12}
\left[\frac{\lambda_a}{2}+T{\rm ln}\left(1+e^{-\lambda_a/T}\right)\right]
+\Omega_e-\Omega_0,
\end{eqnarray}
}
with $\Omega_e=-\mu_Q^4/12\pi^2-\mu_Q^2T^2/6-7\pi^2T^4/180$ being the 
thermodynamic potential of ultra-relativistic electrons, where $\mu_Q=-\mu_e$, 
and  $\Omega_0$ is the divergent vacuum contribution which has to be 
subtracted to assure vanishing energy and pressure of the vacuum.
Out of the twelve eigenvalues $\lambda_a$ four belong to the ungapped blue 
quarks and can be determined easily by using textbook methods 
\cite{Kapusta:1989} as $\lambda_{1..4} = E_f(p)\pm \mu_{fb}$, with the 
dispersion relation $E_f(p)=\sqrt{p^2+M_f^2(p)}$ containing the 
dynamical mass function $M_f(p)=m_f+g(p)\phi_f$ for the two flavors $f=u,d$.
We have introduced the chemical potentials for the quarks of unpaired color
$\mu_{ub}= \mu_B/3+2\mu_Q/3-2\mu_8/2$ and $\mu_{db}=\mu_{ub}-\mu_Q$.  
The other eight eigenvalues belong to the red and green quarks which are paired
in the 2SC state and have therfore identical eigenvalue spectra.
It is thus sufficient to determine the four eigenvalues for the red quarks 
by solving a quartic equation similar to the one discussed in 
\cite{Blaschke:2005uj} for the CFL phase.  
The parameter $\alpha$ describes flavor mixing due to instanton induced 
interactions, see \cite{Frank:2003ve,Buballa:2003qv}. 
For the formfactor we used a simple NJL cut off $g(p)=\theta(1-p/\Lambda)$.
The numerical solutions of this section are obtained with the parameters from 
Table 5.2 of \cite{Buballa:2003qv}; current quark mass 
$m_{u,d} = 5.5~ {\rm MeV}$ , coupling constant $G_s\Lambda^2 = 2.319$, cut off 
$\Lambda = 602.3~ {\rm MeV}$ and diquark coupling $G_D=\eta_D G_s$ with  
$\eta_D = 1$ (strong coupling). 
%
\begin{figure}[ht]
\psfig{figure=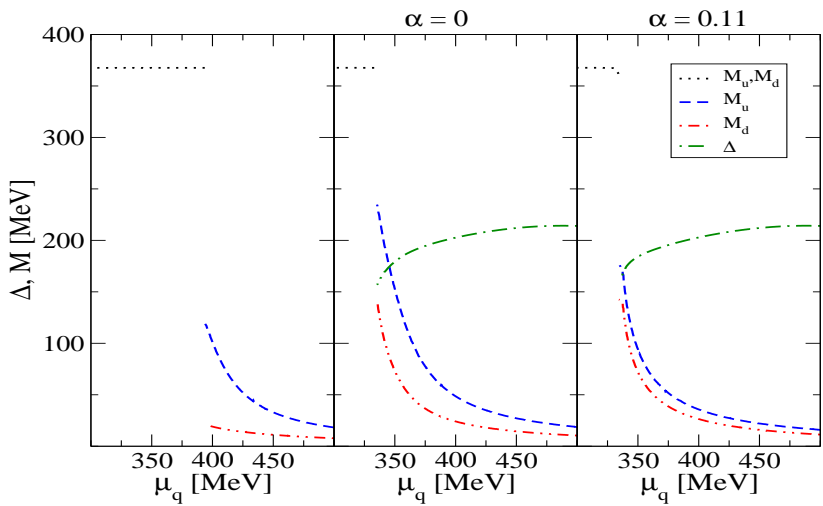,height=7.8cm,width=12.5cm,angle=-90}
\vspace{-0.5cm}
\caption{\small Quark masses for asymmetric matter under charge conservation 
without diquark gap and flavor mixing 
(left panel), with diquark gap and without flavor mixing
(center), and with diquark gap and flavor mixing (right panel).
\label{Fig.1} }
\end{figure}
Fig.~\ref{Fig.1} shows that after the chiral phase transition 
is first order, signalled by a sudden drop of the  
constituent quark masses at the critical chemical potential
$\mu_q^{\rm crit}=340$ MeV.
The region just above the transition,  $\mu_q^{\rm crit}<\mu < 400$ MeV,
relevant for quark stars (or quark cores of neutron stars), shows still
nonperturbatively large quark masses strongly decreasing towards their current 
values with increasing $\mu$. 
In asymmetric quark matter under charge conservation the up quark mass 
is always larger than the down quark mass. 
The magnitude of this mass difference decreases with 
the flavor mixing factor $\alpha$, being maximal for $\alpha = 0$.
Note that the diquark gap $\Delta$ has a similar but smaller mixing effect 
on the quark flavors.     
}
\section{Direct URCA emissivities}
{
The resulting neutrino emissivity from the direct URCA reactions 
\begin{eqnarray}
d &\rightarrow& u + e^- + \bar{\nu}_e~,~~ u + e^- \rightarrow d + \nu_e 
\end{eqnarray}
has been calculated in \cite{Iwamoto:1982} and can be expressed as   
\begin{equation}
\label{urca}
\varepsilon_{\nu} \simeq 
\frac{914~\pi}{1680}G^2\cos^2\theta_c~ p_{F,u}~ p_{F,e}~ p_{F,d}~ T^6~
\left(\frac{1}{3}+\frac{2}{3}\eta\right)~ \theta_{ue}^2. 
\end{equation}
Here $\theta_c$ is the Cabibbo angle, $G$ the weak coupling constant 
and $\theta_{ue}$ is the angle between the up-quark and electron 
momenta obtained from the condition of momentum conservation in the matrix 
element, see Fig. \ref{Fig.2} . 
We take into account that in the 2SC phase $2/3$ of the quarks are paired and 
thus the corresponding direct URCA emissivity is suppressed by a factor 
$\eta =\left(C(\zeta)+\exp[(\Delta-0.5\mu_e)/T]\right)^{-1}$ taken from Ref.
\cite{Jaikumar:2005hy}, where we choose $C(\zeta)=1$. 
For the late cooling stage with temperatures $T\ll 1$ MeV, neutrinos are 
untrapped and their effects on momentum conservation and chemical potentials 
can be neglected.
\begin{figure}[ht]
\hspace{2cm}\psfig{figure=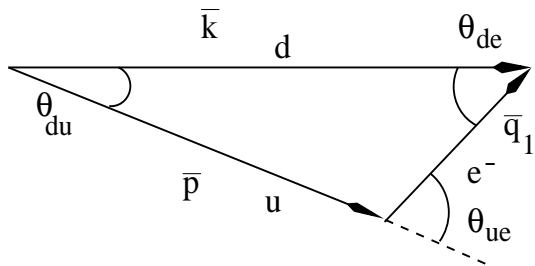,height=3.5cm,width=7cm,angle=0}
\caption{\small Triangle of momentum conservation for the URCA process 
\label{Fig.2}}
\end{figure}
Trigonometric relations can be used to find an appropriate analytical 
expression for the momentum conservation from Fig. \ref{Fig.2}, which in 
lowest order of $\theta_{de}$ is given by
\begin{equation}
\label{eq0}
p_{F,d}-p_{F,u}-p_{F,e}\simeq - \frac{1}{2} p_{F,e}~\theta_{de}^2 ~.
\end{equation}
For small angles holds $\theta_{de} \simeq \theta_{ue}$, so that an expression 
for the matrix element of the direct URCA process can be derived.
Following Iwamoto \cite{Iwamoto:1982} we take into account either quark-quark 
interactions to lowest order in the strong coupling constant $\alpha_s$ 
(\ref{eq1}) or the effect of finite  masses (\ref{eq2})  
\begin{eqnarray}
\mu_i &=& p_{F,i}\left(1+\frac{2}{3\pi}\alpha_s\right) ~,~~~{i=u,d} 
\label{eq1}\\
\mu_i &\simeq& 
p_{F,i}\left[1+\frac{1}{2}\left(\frac{m_i}{p_{F,i}}\right)^2\right]~, 
~~{i=u,d,e}~~. 
\label{eq2}
\end{eqnarray}
From Eqs. (\ref{eq0})-(\ref{eq2}) with the $\beta$-equilibrium condition 
$\mu_d = \mu_u + \mu_e$ an expression for the angle  $\theta_{de}$ determining
the emissivity  Eq. (\ref{urca}) can be found
similar terms for Eq. (\ref{eq0})
\begin{eqnarray}\label{betaeq}
\theta_{de}^2 \simeq \left\{ \begin{array}{cl} 
  \frac{4}{3\pi}\alpha_s \\[0.2cm]
\frac{m_d^2}{p_{F,e}p_{F,d}}
\left[1-\left(\frac{m_u}{m_d}\right)^2\left(\frac{p_{F,d}}{p_{F,u}}\right)
-\left(\frac{m_e}{m_d}\right)^2\left(\frac{p_{F,d}}{p_{F,e}}\right)\right]
\\
\end{array}\right.~.
\end{eqnarray}
\begin{figure}[ht]
\psfig{figure=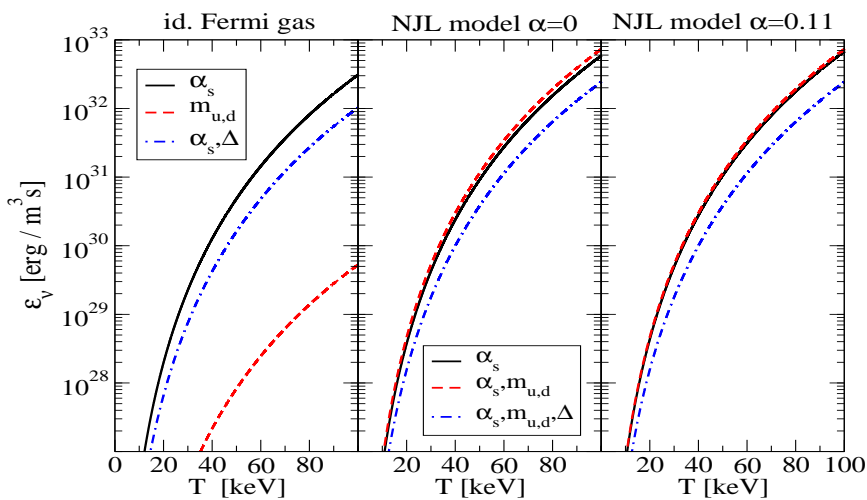,height=7.8cm,width=12.5cm,angle=-90}
\caption{\small Emissivities calculated for parameters obtained from an ideal 
Fermi gas model with the quark current masses  $m_u=5~ {\rm MeV}$ and 
$m_d=9~ {\rm MeV}$ (left panel), vs. a relativistic quark model 
under $\beta$-equilibrium  and charge conservation without flavor mixing 
(center) where $m_u=98~ {\rm MeV}$ and $m_d=45~ {\rm MeV}$  and with flavor 
mixing $\alpha=0.11$  (right panel), where  $m_u=62~ {\rm MeV}$ and 
$m_d=46~ {\rm MeV}$.
The diquark Gap is in all figures $\Delta \simeq 192~{\rm MeV}$.
The solid curves correspond to emissivities via quark-quark interaction, 
dashed curves to the mass effect and dash-dotted curves show the influence 
of a diquark gap. The baryon chemical potential is fixed to 1.1 GeV.
\label{Fig.3}}
\end{figure}
The mass effect for current quark masses is negligible in comparison to the 
perturbative quark-quark interaction effect, as stated in \cite{Iwamoto:1982} 
(see left panel of Fig.~\ref{Fig.3}). 
However, this statement is restricted to the region of the QCD phase diagram, 
where the perturbative treatment is possible at all and the quark masses are 
of the order of the current values $m_{u,d} \sim 5-9~ {\rm MeV}$. 
If $m_u > m_d$ the momentum conservation is violated and the URCA process via 
the mass effect does not work, despite the large quark masses. 
The influence of the quark masses on the Fermi momentum of up and down 
quarks in a relativistic chiral quark model together with the perturbative 
one-gluon exchange (OGE) interactions results in slightly higher emissivities 
compared to the purely perturbative ones. 
Emissivities in superconducting quark matter are suppressed by the  
factor $\eta$, leading to lower emissivities, cf. Fig. \ref{Fig.3}. 
}

\section{Cooling}
{
In order to qualitatively estimate the consequences of the emissivity effects 
discusses above on the cooling behavior of compact stars, we consider the 
simplified model of a homogeneous quark star.
The temperature-age behavior is obtained by inverting the solution of the 
cooling equation 
\begin{equation}
t-t_0=-\int\limits_{T(t_0)}^{T(t)}\sum\limits_{i,j}\frac{ C_v^i(T)}{ L_j(T)}~
{\rm d}T;~~~~~~i={\rm quark},e^-,\gamma,{\rm gluon};~~~~~j=\nu,\gamma.
\end{equation} 
The contributions of the different species to the specific heat and the photon contribution to the star luminosity can be found in \cite{Blaschke:2000dy}. 
\begin{figure}[ht]
\hspace{1cm}\psfig{figure=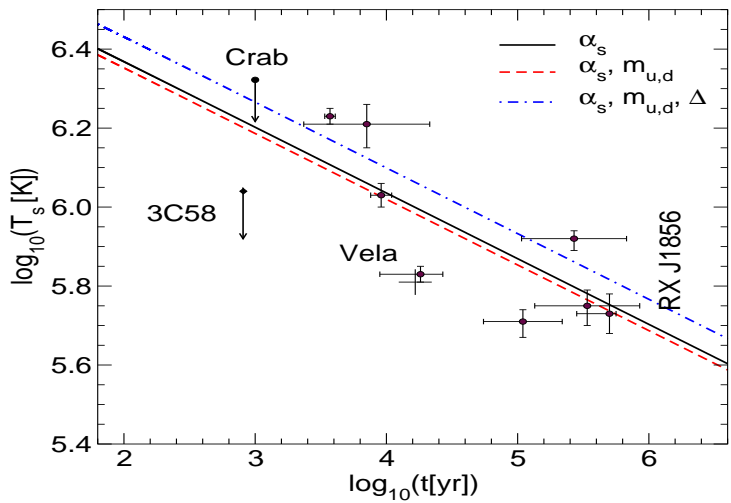,height=7.2cm,width=12cm,angle=-90}
\vspace{-0.5cm}
\caption{\small Cooling curves for the corresponding emissivities from Fig. 
\ref{Fig.3} without flavor mixing (center) for a homogeneous quark star 
($\mu_B=1.1~{\rm GeV}$)  with a radius of $9~{\rm km}$ and a mass 
$1.2~{\rm M}_{\odot}$. .
\label{Fig.4} }
\end{figure}
The double logarithmic plot Fig. \ref{Fig.4} shows the surface temperature 
$T_s$ of the star vs. its age. 
The relation between crust and surface temperatures is taken from the Tsuruta 
law $T_s=(10~ T_c)^{2/3}$.
The cooling curves are sensitive to the quark masses and the diquark gap 
supression. 
Note that the configuration of a homogeneous quark star without hadronic shell 
is quite unrealistic and the comparison with observational data is only made 
to give an orientation about the relations in a T-t plot.      
A realisitic treatment of quark star cooling including heat flux, as well as 
neutrino and photon transport with temperature and density profiles can alter 
these results strongly. 
}

\section{Conclusion}
{ Within our study of the classical URCA process due to perturbative 
OGE interactions we find that constituent quark masses and chemical 
potentials obtained from a NJL type relativistic chiral quark model lead to 
emissivities comparable to the results from the ideal Fermi gas. 
Neutrino emissivities due to the quark masses can only occur if $m_d > m_u$.
Since in the chiral quark model under charge conservation the selfconsistent 
masses obey $m_d < m_u$, momentum conservation is violated and the URCA 
process can not occur. 
In hybrid stars (neutron stars with a quark matter core) or model quark stars 
we can proceed from the assumption that the quark matter close to the phase 
transition is in a non-perturbative density region rather than in a 
perturbative one.
It is questionable whether OGE interactions are applicable
in the non-perturbative density regime close to the chiral phase transition.
Therefore, non-perturbative effects like, e.g., $\rho$ and $\omega$ meson 
exchanges could be possible candidates to replace the perturbative OGE
interaction in the non-perturbative density regime.
}

\end{document}